\documentclass[%
 aip,
 jmp,%
 amsmath,amssymb,
preprint,%
]{revtex4-1}
\usepackage{graphicx}% Include figure files
\usepackage{dcolumn}% Align table columns on decimal point
\usepackage{bm}% bold math
\usepackage{hyperref}% add hypertext capabilities
\input xy
\xyoption{all}

\newtheorem{definition}{Definition}[section]
\newtheorem{lemma}{Lemma}[section]
\newtheorem{proposition}{Proposition}[section]

\begin{document}

\preprint{CIPMA-MPA/0012/2013}

\title[JOURNAL OF MATHEMATICAL PHYSICS]{ P\"oschl-Teller Hamiltonian:  Gazeau-Klauder type coherent states, related statistics and geometry }

\author{Mahouton Norbert Hounkonnou}
\email{norbert.hounkonnou@cipma.uac.bj}

\author{Sama  Arjika}
\email{rjksama2008@gmail.com}

\author{Ezinvi Baloitcha}
\email{ezinvi.baloitcha@cipma.uac.bj}

\affiliation{University of Abomey-Calavi,\\
International Chair in Mathematical Physics and Applications\\
(ICMPA--UNESCO Chair), 072 B.P. 50  Cotonou, Republic of Benin}

\date{\today}

\begin{abstract}
 This work
 mainly addresses a construction of Gazeau-Klauder type coherent states for a  P\"oschl-Teller model. Relevant characteristics are investigated.   Induced geometry and statistics 
are   studied. Then the Berezin - Klauder - Toeplitz quantization of the classical phase space observables is presented.
\end{abstract}

\maketitle
%\tableofcontents

\section{Introduction}
The search for exactly solvable models remains in the core of today research interest in  quantum mechanics.  
A reference list of exactly solvable one-dimensional problems (harmonic oscillator, Coulomb, Morse,
P\"oschl-Teller potentials, etc.)  obtained by an algebraic procedure, namely by a differential operator factorization
methods \cite{infeld}, can be found in \cite{berg} and references therein. This technique, introduced long ago by Schr\"odinger 
\cite{infeld}, was analyzed in depth by Infeld and Hull \cite{ih51}, 
who made an exhaustive classification of factorizable potentials. It was reproduced rather recently in supersymmetric
quantum mechanics (SUSY QM) approach \cite{cooper}
initiated by Witten \cite{wit} 
 and was immediately applied to the hydrogen 
potential \cite{fe84}. This approach gave many new exactly
solvable potentials which were obtained as superpartners of known  exactly solvable models. 
Later 
on, it was  noticed by Witten the possibility of arranging the Schr\"odinger's 
Hamiltonians into isospectral pairs  called {\it supersymmetric partners} \cite{wit}. 
The resulting  supersymmetric quantum mechanics  revived the 
study of exactly solvable Hamiltonians\cite{uh83}.

SUSY QM
is also used for the description
of hidden symmetries of various atomic and nuclear physical systems \cite{gend}.
Besides, it provides a theoretical
laboratory for the investigation of algebraic and dynamical problems in
supersymmetric  field theory.
The simplified setting of SUSY helps to analyze
the  problem of dynamical SUSY breaking at full length
and to examine the validity of the Witten index criterion \cite{wit}.
In \cite{hidden}, it was shown that the reflectionless P\"oschl-Teller system possesses a
hidden bosonized nonlinear (higher order) supersymmetry. This observation
was developed further in \cite{franc1,hidden2}, where it was found that due to a hidden
nonlinear SUSY, the usually super-extended systems possess a much more reach structure than it is 
usually thought. In \cite{franc3}, a nonlinear SUSY of reflectionless PT systems was 
explained using the ideas of AdS/CFT holography and Aharonov-Bohm effect. Exotic nonlinear 
(higher order) SUSY in pairs of mutually shifted reflectionless 
PT systems was studied in \cite{mik1}. Its 
relation to kink-antikink crystal appearing in Gross-Neveu model was 
studied in \cite{mik2} and \cite{adrian1}.  Extension of such class of the 
systems with higher order supersymmetry
for the case of PT-symmetry was investigated recently in \cite{hidden1} and \cite{hidden3}.
Besides, in a recent paper \cite{adrian2,Arancibia}, an exotic 
nonlinear (higher order) supersymmetry was investigated in a
much more general class of soliton systems.
Recently, in \cite{berg}, Bergeron et {\it al}
 developed the mathematical aspects  
that have been left apart in \cite{gazeau}
(proof of the resolution
of unity, detailed calculations of quantized version of classical observables 
and mathematical study of the resulting operators: problems of domains, 
self-adjointness or self-adjoint extensions). Some additional questions 
as asymptotic behavior were also studied. Moreover,  
extensions were discussed to a larger class of P\"oschl-Teller potentials.

This paper is organized as follows. In Section 2, we recall known results and some intertwining relations.
  In Section 3,  we build the 
 associated  Gazeau-Klauder coherent states (GKCSs).
  Their main mathematical properties, i.e., the orthogonality, the normalizability, the continuity in the labels and the resolution of the identity are investigated. Quantum statistics and geometry   of these states are studied in  Section 4. Section 5 is devoted to the Berezin - Klauder - Toeplitz quantization of classical phase space observables. We end with a conclusion in  Section 6.

\section{The P\"oschl-Teller Hamiltonian and SUSY-QM formalism}
In this section, for the clarity of our development, we   briefly recall  the 
 P\"oschl-Teller Hamiltonian  model presented in  \cite{berg} and summarize main results on the eigenvalue problem and supersymmetry factorization method of the time independent Schr\"odinger equation. 
\subsection{The model}
The physical system is described by the Hamiltonian \cite{berg}
\begin{eqnarray}
\label{ham}
{\bf H_{\nu,\beta}}\phi:=\Big(-\frac{\hbar^2}{2M}\frac{d^2}{dx^2}+V_{\varepsilon_0,\nu,\beta}(x)\Big)
\phi\quad\mbox{for}\quad \phi\in\mathcal{D}_{\bf H_{\nu,\beta}}
\end{eqnarray}
in  a
 suitable  Hilbert space 
 $\mathcal{H}=L^2([0,L],dx)$ endowed with the inner product defined by 
\begin{eqnarray}
(u,v)=\int_{0}^{L}dx\,\bar{u}(x)v(x),\quad u,v\in\mathcal{H},\; [0,L]\subset\mathbb{R}
\end{eqnarray}
where $\bar{u}$ denotes the complex conjugate of $u.$   $M$ is the particle mass and $\mathcal{D}_{\bf H_{\nu,\beta}}$
 is the domain of definition of ${\bf H_{\nu,\beta}}$. 

\begin{eqnarray}
\label{hamk}
V_{\varepsilon_0,\nu,\beta}(x)=\varepsilon_0
\Big(\frac{\nu(\nu+1)}{\sin^2\frac{\pi x}{L}}-2\beta\cot\frac{\pi x}{L}\Big)
\end{eqnarray}
is the P\"oschl-Teller potential;
 $\varepsilon_0$ is some energy scale, $\nu$ and $\beta$ are  dimensionless parameters. 

 The one-dimensional second-order operator ${\bf H_{\nu,\beta}}$ has singularities at the end points $x=0$ and $x=L$  permiting  to choose $\varepsilon_0\geq 0$
and $\nu\geq 0$. Further, since the symmetry $x\rightarrow L-x$ corresponds to the parameter change $\beta\rightarrow -\beta,$ we can choose $\beta\geq 0.$  As assumed in \cite{berg}, we consider the 
energy scale  
$\varepsilon_0$ as the zero point energy of the energy of the infinite well, i.e. $\varepsilon_0=\hbar^2\pi^2/(2ML^2)$ so that the unique free  parameters of the problem remain
$\nu$ and $\beta$  which will be always assumed to be positive.  The case $\beta=0$ corresponds to the symmetric repulsive potentials investigated in \cite{gazeau}, while the case 
$\beta\neq 0$ leads to the Coloumb potential in the limit $L\rightarrow \infty.$ 

Let us define  the  operator ${\bf \mathcal{H}_{\nu,\beta}}$
with the action $-\frac{\hbar^2}{2M}\phi''(x)+\varepsilon_0
\Big(\frac{\nu(\nu+1)}{\sin^2\frac{\pi x}{L}}-2\beta\cot\frac{\pi x}{L}\Big)\phi(x)$   with the domain being the set of smooth functions with a compact support, $\mathcal{C}_0^\infty(0,L)$.
The P\"oschl-Teller potential is in the limit point case at both ends $x=0$ and $x=L$, if the parameter $\nu\geq 1/2$, and in the limit circle case at both ends if $0\leq \nu< 1/2. $ Therefore, the operator ${\bf \mathcal{H}_{\nu,\beta}}$ is essentially  self-adjoint  in the former case. The closure of ${\bf \mathcal{H}_{\nu,\beta}}$ is 
$\overline{\bf \mathcal{H}_{\nu,\beta}}={\bf H_{\nu,\beta}},$ i.e., $\mathcal{D}_{\overline{\bf \mathcal{H}_{\nu,\beta}}}=\mathcal{D}_{\bf H_{\nu,\beta}}$ and its domain coincides with the maximal one,
i.e.,
\begin{eqnarray}
\mathcal{D}_{{\bf H_{\nu,\beta}}}=\Bigg\{\phi\in ac^2(0,L),\; \Big[-\frac{\hbar^2}{2M}\phi''+\varepsilon_0
\Big(\frac{\nu(\nu+1)}{\sin^2\frac{\pi x}{L}}-2\beta\cot\frac{\pi x}{L}\Big)\phi\Big]\in\mathcal{H}\Bigg\},
\end{eqnarray} 
 where $ac^2(0,L)$ denotes the absolutely continuous functions with
absolutely continuous derivatives. As mentioned in \cite{berg}, a function of this domain satisfies  Dirichlet boundary conditions 
and in the range of considered $\nu$, 
the deficiency indices of ${\bf \mathcal{H}_{\nu,\beta}}$ is $(2,2)$ indicating that this operator is no longer
essentially self-adjoint but has  a two-parameter family of self-adjoint extensions indeed. As in \cite{berg}, we will restrict only
 to the extension described by Dirichlet boundary conditions, i.e., 
$$\mathcal{D}_{{\bf H_{\nu,\beta}}}=\Bigg\{\phi\in ac^2(0,L), \mid \phi(0)= \phi(L)=0,\; \Big[-\frac{\hbar^2}{2M}\phi''+\varepsilon_0
\Big(\frac{\nu(\nu+1)}{\sin^2\frac{\pi x}{L}}-2\beta\cot\frac{\pi x}{L}\Big)\phi\Big]\in\mathcal{H}\Bigg\}.$$ 
 $\mathcal{D}_{{\bf H_{\nu,\beta}}}$ is  dense in
$\mathcal{H}$ since $H^{2,2}(0, L)\supset\mathcal{D}_{{\bf H_{\nu,\beta}}} $ 
and  ${\bf H_{\nu,\beta}}$ is self-adjoint,
 where  $H^{m,n}(0, L)$ is the Sobolev space of indice $(m,n)$ \cite{sob}. Later on, we use the dense domain 
\begin{eqnarray}
\mathcal{D}_H=\Big\{\phi\in AC^2(0,L), \,\varepsilon_0
\Big(\frac{\nu(\nu+1)}{\sin^2\frac{\pi x}{L}}-2\beta\cot\frac{\pi x}{L}\Big)\phi\in\mathcal{H}\Big\},
\end{eqnarray}
and $AC(0,L)$ is defined as 
\begin{eqnarray}
AC(0,L)=\Big\{\phi\in ac(0,L):\phi^{'}\in\mathcal{H}\Big\}.
\end{eqnarray}
\subsection{Eigenvalues and eigenfunctions}
The eigenvalues $E_{n}^{(\nu,\beta)}$ and functions $\phi_{n}^{(\nu,\beta)}$  solving the  Sturm-Liouville differential 
equation (\ref{ham}),  i.e.,
${\bf H_{\nu,\beta}}\phi_{n}^{(\nu,\beta)}=E_{n}^{(\nu,\beta)}\phi_{n}^{(\nu,\beta)},$
 are given by \cite{berg}
\begin{eqnarray}
\label{eigen1}
E_{n}^{(\nu,\beta)}=\varepsilon_0\left((n+\nu+1)^2-\frac{\beta^2}{(n+\nu+1)^2}\right)
\end{eqnarray}
and
\begin{eqnarray}
\label{eigen2}
  \phi_{n}^{(\nu,\beta)}(x)&=&K_n^{(\nu,\beta)}\sin^{\nu+n+1}\left(\frac{\pi x}{L}\right)
\exp \left(-\frac{\beta \pi x}{L(\nu+n+1)}\right)P_n^{(a_n,\bar{a}_n)}\Big(i\cot \frac{\pi x}{L}\Big)
\end{eqnarray}
respectively, where $n\in\mathbb{N}$, $a_n=-(n+\nu+1)+i\frac{\beta}{n+\nu+1}$, $P_n^{(\lambda,\eta)}(z)$  
are the Jacobi polynomials \cite{ASK} and 
$K_n^{(\nu,\beta)}$ is a normalization constant giving by
\begin{eqnarray}
\label{prpos}
K_n^{(\nu,\beta)}
=2^{n+\nu+1}L^{-\frac{1}{2}}
\mathcal{T}(n;\nu,\beta)
\mathcal{O}^{-\frac{1}{2}}(n;\nu,\beta)\exp\left\{\frac{\beta\pi}{2(n+\nu+1)}\right\},
\end{eqnarray}
where
\begin{eqnarray}
\label{rem}
\mathcal{O}(n;\nu,\beta)&=&\sum_{k=0}^n   \frac{(-n,-2\nu-n-1)_k}{(-\nu-n-\frac{i\beta}{\nu+n+1})_kk!\Gamma(n+\nu+2-k+\frac{i\beta}{\nu+n+1})}\cr
&\times&\sum_{s=0}^n \frac{(-n,-2\nu-n-1)_s\Gamma(2n+2\nu-s-k+3)}{ (-\nu-n+\frac{i\beta}{\nu+n+1})_s s! \Gamma(n+\nu+2-s-\frac{i\beta}{\nu+n+1})}
\end{eqnarray}
and
\begin{eqnarray}
\label{remo}
\mathcal{T}(n;\nu,\beta)=n!\Big|\Big(-n-\nu+\frac{i\beta}{n+\nu+1}\Big)_n\Big|^{-1}.
\end{eqnarray}
For details on the $K_n^{(\nu,\beta)},$ see Appendix A.

For $n=0$, one can retrieve                                                                                                                                                                                                                                                                                                                                                                                                                                                                                                                                                                                                                                                                                                                                                                                                                                                                                                                                                                                                                                     
                                                                                                                                                                                        
\begin{eqnarray}
\label{eigen2e}
\phi_0^{(\nu,\beta)}(x)=\frac{2^{\nu+1}}{\sqrt{L\times\Gamma(2\nu+3)}}
\sin^{\nu+1}\left(\frac{\pi x}{L}\right)
\exp \left\{\frac{\beta \pi }{\nu+1}\Big(\frac{1}{2}-\frac{x}{L}\Big)\right\}.
\end{eqnarray}
\subsection{Factorization method, shape invariance of the
 P\"oschl-Teller Hamiltonian and intertwining relations}
We assume the 
ground state $\phi_0^{(\nu,\beta)}$
and the energy  $E_0^{(\nu,\beta)}$ are  known. By using  a  Darboux factorization  method of the Hamiltonian \cite{ih51,cooper,david2}, one can 
 define the differential operators 
$A_{\nu,\beta},\; A_{\nu,\beta}^\dag $  factorizing 
the P\"oschl-Teller  Hamiltonian 
${\bf H_{\nu,\beta}}$  (\ref{ham}) and the associated
 super potential  $W_{\nu,\beta}$ 
as follows
\begin{eqnarray}
\label{fact}
{\bf H_{\nu,\beta}}:=\frac{1}{2M}A_{\nu,\beta}^\dag A_{\nu,\beta}+E_0^{(\nu,\beta)},
\end{eqnarray}
where the differential operators $A_{\nu,\beta}$ and $A_{\nu,\beta}^\dag$ are defined as
\begin{eqnarray}
\label{facto}
A_{\nu,\beta}:=\hbar\frac{d}{dx}+W_{\nu,\beta}(x),\quad 
A_{\nu,\beta}^\dag :=-\hbar\frac{d}{dx}+W_{\nu,\beta}(x),
\end{eqnarray}
acting in  the domains
\begin{eqnarray}
\label{maine}
\mathcal{D}_{A_{\nu,\beta}}&=&\{\phi\in ac(0,L)| \;(\hbar \phi'+W_{\nu,\beta}\phi)\in\mathcal{H}\},
\end{eqnarray}
and 
\begin{eqnarray}
 \mathcal{D}_{A_{\nu,\beta}^\dag}=\left\{\phi\in ac(0,L)|\;\exists \;\tilde{\phi}\in \mathcal{H}:
[\hbar\psi(x)\phi(x)]_0^L=0,\;\langle A_{\nu,\beta} \psi,\phi \rangle= \langle
 \psi,\tilde{\phi }\rangle,\;\forall\;\psi\in\mathcal{D}_{A_{\nu,\beta}}\right\},\nonumber
\end{eqnarray}
where $A_{\nu,\beta}^\dag \phi=\tilde{\phi}$. 
The operator $A_{\nu,\beta}^\dag$ is the adjoint of $A_{\nu,\beta}.$
 Besides,
 considering their common restriction
\begin{eqnarray}
\label{maine}
\mathcal{D}_{A}=\{\phi\in AC(0,L)| \;W_{\nu,\beta}\phi\in\mathcal{H}\},
\end{eqnarray}
we have
$\overline{A_{\nu,\beta}\upharpoonright\mathcal{D}_{A}}=A_{\nu,\beta}$
and $\overline{A_{\nu,\beta}^\dag\upharpoonright\mathcal{D}_{A}}=A_{\nu,\beta}^\dag$ (for more details, see \cite{berg}).
The super-potential $W_{\nu,\beta}$ is 
given by
\begin{eqnarray}
\label{factoz}
 W_{\nu,\beta}(x):=-\hbar\frac{[\phi_0^{(\nu,\beta)}(x)]'}{\phi_0^{(\nu,\beta)}(x)}=-\frac{\pi\hbar}{L}\Big( (\nu+1)\cot\frac{\pi x}{L}-\frac{\beta}{\nu+1}\Big).
\end{eqnarray}
The superpartner Hamiltonian $
{\bf H}_{\nu,\beta}^{(1)}$ of  ${\bf H_{\nu,\beta}}$ is obtained by permuting the operators $A_{\nu,\beta}^\dag$ and 
$A_{\nu,\beta} $ and we get  
\begin{eqnarray}
\label{facte}
{\bf H}^{(1)}_{\nu,\beta}:=\frac{1}{2M}A_{\nu,\beta}  A_{\nu,\beta} ^\dag+E_0^{(\nu,\beta)}=-\frac{\hbar^2}{2M}\frac{d^2}{dx^2}+
V_{\nu,\beta}^{(1)}(x),
\end{eqnarray}
where the partnerpotential $V_{\nu,\beta}^{(1)}$ of $V_{\varepsilon_0, \nu,\beta}$ is defined by the relation
\begin{eqnarray}
\label{sama:poten}
V_{\nu,\beta}^{(1)}(x)&:=&\frac{1}{2M}\Big(W^2_{\nu,\beta}(x)+W'_{\nu,\beta}(x)\Big)+E_0^{(\nu,\beta)}.
\end{eqnarray}
Performing (\ref{sama:poten}), we arrive at the following relation
\begin{eqnarray}
V_{\nu,\beta}^{(1)}(x)\equiv V_{\varepsilon_0,\nu+1,\beta}(x).
\end{eqnarray}
Therefore, the superpartner Hamiltonian (\ref{facte}) becomes
\begin{eqnarray}
\label{sama:sup}
{\bf H}^{(1)}_{\nu,\beta}\equiv{\bf H}_{\nu+1,\beta},
\end{eqnarray}
This relation specifies that P\"oschl-Teller Hamiltonians are shape invariant, i.e.,
\begin{eqnarray}
A_{\nu,\beta}  A_{\nu,\beta} ^{ \dag}=A_{\nu+1,\beta} ^{ \dag} A_{\nu+1,\beta} +2M(E_0^{(\nu+1,\beta)}-E_0^{(\nu,\beta)}).
\end{eqnarray}
In
the equation
\begin{eqnarray}
\label{eqnn}
{\bf H}_{\nu,\beta}^{(1)}|\phi_n^{(1,\nu,\beta)}\rangle=E_n^{(1,\nu,\beta)}|\phi_n^{(1,\nu,\beta)}\rangle,
\end{eqnarray}
the eigenfunction  $|\phi_n^{(1,\nu,\beta)}\rangle$ and the eigenvalue $E_n^{(1,\nu,\beta)}$  of ${\bf  H}_{\nu,\beta}^{(1)}$ are
related to those of ${\bf H}_{\nu,\beta}$, i.e., $E_n^{(1,\nu,\beta)}=E_n^{(\nu+1,\beta)}:=E_{n+1}^{(\nu,\beta)}$ and 
\begin{eqnarray}
\label{sama:superpo}
  |\phi_n^{(\nu+1,\beta)}\rangle=
\frac{A_{\nu,\beta}|\phi_{n+1}^{(\nu,\beta)}\rangle}{\sqrt{2M(E_{n+1}^{(\nu,\beta)}-E_0^{(\nu,\beta)})}}
\qquad |\phi_{n+1}^{(\nu,\beta)}\rangle=
\frac{A_{\nu,\beta}^\dag|\phi_{n}^{(\nu+1,\beta)}\rangle}{\sqrt{2M(E_{n+1}^{(\nu,\beta)}-E_0^{(\nu,\beta)})}}
\end{eqnarray} 
 satisfying
\begin{eqnarray}
 \langle\phi_m^{(\nu,\beta)} |\phi_n^{(\nu,\beta)}\rangle=\delta_{mn}\quad  \mbox{ and } \quad
\sum_{n=0}^\infty |\phi_n^{(\nu,\beta)}\rangle \langle \phi_n^{(\nu,\beta)}|=\mathbb{I}.
\end{eqnarray}
Now let us introduce the positive sequence $\eta_n^{(\nu,\beta)}=
\varepsilon_0^{-1}\left( E_n^{(\nu,\beta)}-
E_0^{(\nu,\beta)} \right) $. Then, the 
representations
of the operators  $A_{\nu,\beta}$ and
$A_{\nu,\beta}^{\dag}$  are  given by
\begin{eqnarray}
\label{sama:dec}
 A_{\nu,\beta}=\sqrt{2M \varepsilon_0} 
\sum_{n=0}^\infty \sqrt{\eta_{n+1}^{(\nu,
\beta)}}\;|{\phi}_n^{(\nu+1,\beta)}
\rangle\langle\phi_{n+1}^{(\nu,\beta)}|
\end{eqnarray}
and
\begin{eqnarray}
\label{sama:dec1}
&&A^\dag_{\nu,\beta} = \sqrt{2M \varepsilon_0} 
\sum_{n=0}^\infty \sqrt{\eta_{n+1}^{(\nu,\beta)}}\;|\phi_{n+1}^{(\nu,
\beta)}\rangle\langle \phi_n^{(\nu+1,\beta)}|,
\end{eqnarray}
respectively.
Any eigenstate $\phi_n^{(\nu,\beta)}(x)$ ($n=1,2,\cdots$) of 
${\bf H_{\nu,\beta}}$ may then be constructed 
from the ground state $\phi_0^{(\nu+n,\beta)}(x)$
 through the repeated 
application of $A_{\nu+k,\beta}^\dag,\;k=0,1,2,\cdots,n-1$ operators 
defined in terms of the superpotential, i.e.,
\begin{equation}
\label{sama:func}
 \phi_n^{(\nu,\beta)}(x)
\propto A_{\nu,\beta}^\dag 
A_{\nu+1,\beta}^\dag \ldots A_{\nu+n-1,\beta}^\dag
\phi_0^{(\nu+n,\beta)}(x).
\end{equation}
The operators  $A_{\nu,\beta}$ and 
$A_{\nu,\beta}^\dag$ do not commute 
with the  P\"oschl-Teller  Hamiltonian 
${\bf H}_{\nu,\beta}$, but satisfy the intertwining relations
\begin{eqnarray}
\label{46}
{\bf H_{\nu,\beta}}A_{\nu,\beta}^\dag =
A_{\nu,\beta}^\dag {\bf H}_{\nu+1,\beta},
\quad A_{\nu,\beta}{\bf H_{\nu,\beta}} =
{\bf H}_{\nu+1,\beta}A_{\nu,\beta}.
\end{eqnarray}
More generally,
\begin{eqnarray}
\label{sama:relas}
{\bf H_{\nu,\beta}}B_n=B_n{\bf H}_{\nu+n,\beta},\quad 
B_n^\dag{\bf H_{\nu,\beta}}={\bf H}_{\nu+n,\beta}B_n^\dag,
\end{eqnarray}
where $B_n$ and $B_n^\dag$ are operators of degree $n$, i.e.,
\begin{eqnarray}
\label{seu}
B_n:=A_{\nu,\beta}^\dag A_{\nu+1,
\beta}^\dag\ldots A_{\nu+n-1,\beta}^\dag,\quad 
B_n^\dag:=A_{\nu+n-1,\beta}\ldots 
A_{\nu+1,\beta}A_{\nu,\beta}. 
\end{eqnarray}
Therefore, for any positive integers $n, m,$  
the following result holds:
\begin{eqnarray}
\label{5dp1ed}
B_m^\dag B_n=
(2M)^{n}{_{m}\Lambda}_{n,\nu,\beta}
\prod_{k=0}^{n-1}\Big({\bf H}_{\nu+n,\beta}-
E_0^{(\nu+k,\beta)}\Big) 
\end{eqnarray}
if $n < m,$
\begin{eqnarray}
\label{5dp1eod}
B_m^\dag B_n=
(2M)^{m}\prod_{k=0}^{m-1}\Big({\bf H}_{\nu+m,\beta}-E_0^{(\nu+k,\beta)}\Big)\;
{\Theta}_{n,\nu,\beta}^m
\end{eqnarray}
if $n > m$, where the operators $ {_{m}\Lambda}_{n,\nu,\beta}$ 
and ${\Theta}_{n,\nu,\beta}^m$  are given by
\begin{eqnarray}
\label{5dedp}
  {_{m}\Lambda}_{n,\nu,\beta}:=A_{\nu+m-1,
\beta}A_{\nu+m-2,\beta}\ldots A_{\nu+n,\beta},\;\;
{\Theta}_{n,\nu,\beta}^m:=A_{\nu+m,\beta}^\dag A_{\nu+m+1,\beta}^\dag\ldots A_{\nu+n-1,\beta}^\dag.
\end{eqnarray}
In particular, for $n = m$, we have
\begin{eqnarray}
\label{5dp}
  B_n B_n^\dag=(2M)^{n}\prod_{k=0}^{n-1}
\Big({\bf H_{\nu,\beta}}-E_0^{(\nu+k,\beta)}\Big),\quad
B_n^\dag B_n=(2M)^{n}\prod_{k=0}^{n-1}
\Big({\bf H}_{\nu+n,\beta}-E_0^{(\nu+k,\beta)}\Big).
\end{eqnarray}

The operators ${_{m}\Lambda}_{n,\nu,\beta}$ and ${\Theta}_{n,\nu,\beta}^m$ satisfy the following identities
\begin{eqnarray}
\label{4etttp7}
&&{_{m}\Lambda}_{n,\nu,\beta}^\dag\,{_{m}\Lambda}_{n,\nu,\beta}=
(2M)^{m-n}\prod_{k=n}^{m-1}\Big({\bf H}_{\nu+n,\beta}-E_0^{(\nu+k,\beta)}\Big),\\
\label{4ettt7z}
&&{\Theta}_{n,\nu,\beta}^m\,{\Theta}_{n,\nu,\beta}^{m \dag}=
(2M)^{n-m}\prod_{k=m}^{n-1}\Big({\bf H}_{\nu+m,\beta}-E_0^{(\nu+k,\beta)}\Big).
\end{eqnarray}
Indeed, from (\ref{sama:func}) 
and (\ref{seu}), one 
can see that the actions of  the operators 
$B_n^\dag$ and  $B_n$  on the normalized 
eigenfunctions $\phi_n^{( \nu,\beta)}$ and  
$\phi_0^{(\nu+n,\beta)}$  of 
$\;{\bf H}_{\nu,\beta}$
are given by
\begin{eqnarray}
\label{48}
B_n^\dag \phi_n^{(\nu,\beta)}(x)= 
\big(\pi\hbar L^{-1}\big)^n\mathcal{M}^{1/2}(n;\nu,\beta)\,
\phi_0^{(\nu+n,\beta)}(x)
\end{eqnarray}
and
\begin{eqnarray}
\label{cr}
B_n  \phi_0^{(\nu+n,\beta)}(x)= \big(\pi\hbar 
L^{-1}\big)^n\mathcal{M}^{1/2}(n;\nu,\beta)\,\phi_n^{(\nu,\beta)}(x),
\end{eqnarray}
respectively, 
where 
\begin{eqnarray}
 \mathcal{M}(n;\nu,\beta)
=\prod_{k=0}^{n-1}\left(E_n^{(\nu,\beta)}-E_k^{(\nu,\beta)}\right).
\end{eqnarray}

The mean values of the operators $B_n B_n^\dag$,  $B_n^\dag 
B_n$, ${_{m}\Lambda}_{n,\nu,\beta}^\dag\,{_{m}\Lambda}_{n,\nu,\beta}$ and   ${\Theta}_{n,\nu,\beta}^m\,{\Theta}_{n,\nu,\beta}^{m \dag}$ in the states $ |\phi_n^{(\nu,\beta)}\rangle$ are derived    
by using  (\ref{5dp}). We obtain  
\begin{eqnarray}
\label{4d8}
   \langle B_n B_n^\dag  \rangle_{
\phi_n^{(\nu,\beta)}}=(\pi\hbar 
L^{-1})^{2n}\mathcal{M}(n;\nu,\beta), \quad
\langle B_n^\dag B_n \rangle_{
\phi_n^{(\nu,\beta)}}=(\pi\hbar L^{-1})^{2n}
\mathcal{T}(n;\nu,\beta),
\end{eqnarray}
\begin{eqnarray}
\label{lave}
\langle{_{m}\Lambda}_{n,\nu,\beta}^\dag\,{_{m}\Lambda}_{n,\nu,\beta}\rangle_{
\phi_n^{(\nu,\beta)}}=(\hbar\pi L^{-1})^{2m-2n}\;\prod_{k=n}^{m-1}\left(E_{2n}^{(\nu,\beta)}-E_k^{(\nu,\beta)}\right)
%\mathcal{N}(m,n;\nu,\beta),
\quad n<m,
\end{eqnarray}
and
\begin{eqnarray}
\label{des}
\langle{\Theta}_{n,\nu,\beta}^m\,{\Theta}_{n,\nu,\beta}^{m \dag}\rangle_{
\phi_n^{(\nu,\beta)}}=(\hbar\pi L^{-1})^{2n-2m}\prod_{k=m}^{n-1}\left(E_{n+m}^{(\nu,\beta)}-E_k^{(\nu,\beta)}\right)\;%\mathcal{N}(n,m;\nu,\beta),
\quad n> m,
\end{eqnarray}
where 
\begin{eqnarray}
  \mathcal{T} (n;\nu,\beta)
=\prod_{k=0}^{n-1}\left(E_{2n}^{(\nu,\beta)}-E_k^{(\nu,\beta)}\right)
\end{eqnarray}
and
\begin{eqnarray}
\langle  A_{\nu,\beta} \rangle_{
\phi_n^{(\nu,\beta)}}:=\int_0^L dx\;
\overline{\phi_{n}^{(\nu,\beta)}(x)}
A_{\nu,\beta}\phi_{n}^{(\nu,\beta)}(x).
\end{eqnarray}

In the sequel, the parameter $\beta=0$.
\section{Gazeau-Klauder  coherent states (GKCSs)}
Without loss of generality, let us consider the P\"oschl-Teller  Hamiltonian  (\ref{ham}) with the parameter  $\beta=0$.  The resulting  one-dimensional quantum
mechanical system has a  
finite or 
infinite number of discrete energy levels 
$\mathcal{E}_n^{(\nu)},\;n\,\in\,\mathbb{N},$
\begin{eqnarray}
\label{sama:eigen1}
\mathcal {E}_n^{(\nu)}:=2M(E_n^{(\nu)}-E_0^{(\nu)})
\end{eqnarray}
  chosen
by adjusting the constant
part of the P\"oschl-Teller  Hamiltonian 
${\bf H}_{\nu,0}$, and satisfying $\mathcal{E}_n^{(\nu)}<
\mathcal{E}_{n+1}^{(\nu)}.
$ 
 Then  the resulting  positive definite  
Hamiltonian $\mathcal {H}_{\nu}$   is expressed in a 
factorized form as follows
\begin{eqnarray}
\label{sama:hamil}
{\mathcal H}_{\nu}:=A_{\nu}^\dag A_{\nu},
\end{eqnarray}
where  the differential operators 
$A_{\nu}$ and $A_{\nu}^\dag$ are defined 
as
\begin{eqnarray}
\label{sama:eigen22}
A_{\nu}:=A_{\nu,0} \quad \mbox{ and }\quad 
A_{\nu}^\dag:=A_{\nu,0}^\dag
\end{eqnarray}
and 
%According to (\ref{sama:dec}) and  (\ref{sama:dec1}), 
their
representations are  given, respectively, by
\begin{eqnarray}
&&A_{\nu}=
\sum_{n=0}^\infty \sqrt{\mathcal {E}_{n+1}^{(\nu)}}\;|{\phi}_n^{(\nu+1,0)}
\rangle\langle\phi_{n+1}^{(\nu,0)}|,\\
%\end{equation}
%\begin{equation}
&&A^\dag_{\nu} = 
\sum_{n=0}^\infty \sqrt{\mathcal {E}_{n+1}^{(\nu)}}\;|\phi_{n+1}^{(\nu,
0)}\rangle\langle \phi_n^{(\nu+1,0)}|,
\end{eqnarray}
and 
\begin{equation}
{\mathcal H}_{\nu} = 
\sum_{n=0}^\infty   \mathcal {E}_{n+1}^{(\nu)}\;|\phi_{n+1}^{(\nu,
0)}\rangle\langle \phi_{n+1}^{(\nu,0)}|.
\end{equation}
The eigenvalues  $\mathcal {E}_{n}^{(\nu)}:=\mathcal {E}^{(\nu)}(n)$
 that solve the   eigenvalue problem related to  (\ref{sama:hamil}),  i.e.,
$\mathcal {H}_{\nu}|n\rangle_{\nu}=\mathcal {E}^{(\nu)}(n)
|n\rangle_{\nu}$
 are  given by \cite{gazeau} 
\begin{eqnarray}
\label{sama:eigen1}
\mathcal {E}^{(\nu)}(n)=
2M\,\varepsilon_0\,n(n+2\nu+2).
\end{eqnarray}

Let $\mathcal{F}$ be the Fock space spanned by 
$\{|n\rangle_{\nu},\,n = 0, 1,...,\}$
satisfying the useful conditions
\begin{eqnarray}
 _\nu\langle m|
n\rangle_\nu=\delta_{mn},\qquad \sum_{n=0}^\infty|n\rangle_\nu
\,_\nu\langle n|={\bf 1} 
\end{eqnarray}
 such that
\begin{eqnarray}
\label{sama:eigen2}
 |n\rangle_{\nu}:=Z_{n,\nu}\sin^{\nu+1}\left(\frac{\pi x}{L}\right)
C_n^{\nu+1}\Big(\cos \frac{\pi x}{L}\Big),\quad n\in\mathbb{N},
\end{eqnarray}
where  $C_n^{(\theta)}(z)$  is a Gegenbauer polynomial
\cite{ASK} and $Z_{n,\nu}$ is a normalization constant \cite{gazeau}.
We assume that there exists a real number $\gamma$ such that the actions of the $\gamma-$depending annihilation and creation-like 
operators $_\gamma A_{\nu}$ and $_\gamma A_{\nu}^\dag$ on
(\ref{sama:eigen2}) are given by \cite{Davidj,KinaniD}
\begin{eqnarray}
\label{sama:a}
_\gamma A_{\nu} |n\rangle_{\nu}:= \sqrt{ \mathcal {E}^{(\nu)} (n)}\,e^{i\gamma( \mathcal {E}^{(\nu)} (n)-\, \mathcal{E}^{(\nu)}(n-1))}|n-1\rangle_{\nu}
\end{eqnarray}
and 
\begin{eqnarray}
\label{sama:adag}
_\gamma A_{\nu}^\dag |n\rangle_{\nu}:= \sqrt{ \mathcal {E}^{(\nu)} (n+1)}\,e^{-i\gamma( \mathcal {E}^{(\nu)} (n+1)-\, \mathcal{E}^{(\nu)}(n))} |n+1\rangle_{\nu},
\end{eqnarray}
respectively. Indeed,
\begin{eqnarray}
\label{sama:asas}
_\gamma A_{\nu}\,{ _\gamma }A_{\nu}^\dag|n\rangle_{\nu}=  \mathcal {E}^{(\nu)} (n+1)|n\rangle_{\nu},\quad
_\gamma A_{\nu}^\dag\, {_\gamma }A_{\nu} |n\rangle_{\nu}= 
\mathcal {E}^{(\nu)} (n)|n\rangle_{\nu}.
\end{eqnarray}
Formally, 
\begin{eqnarray}
N|n\rangle_{\nu}=n|n\rangle_{\nu}.
\end{eqnarray}
The relations in (\ref{sama:asas}) lead to 
\begin{eqnarray}
_\gamma A_{\nu}\, {_\gamma }A_{\nu}^\dag=\mathcal {E}^{(\nu)} (N+1),
\quad _\gamma A_{\nu}^\dag\, {_\gamma} A_{\nu} =
\mathcal {E}^{(\nu)} (N)
\end{eqnarray}
and we arrive at the following set of 
non-null commutation relation 
of the  algebra of $\mathcal {H}_{\nu}$
\begin{eqnarray}
\label{sama:algebra}
&&_\gamma A_{\nu}\, {_\gamma }A_{\nu}^\dag-\,{ _\gamma} A_{\nu}^\dag\, {_\gamma} A_{\nu} =f_\nu(N),\; [N,\,{_\gamma} A_{\nu}]=-\,{_\gamma} A_{\nu},\,
[N, \,_\gamma A_{\nu}^\dag]= \,_\gamma A_{\nu}^\dag,\\
&&[\mathcal {H}_{\nu},{_\gamma} A_{\nu}]=-\,{_\gamma} A_{\nu}f_\nu(N-1),\;[\mathcal {H}_{\nu},\,{_\gamma}A_{\nu}]={_\gamma} A_{\nu}^\dag f_\nu(N)
\end{eqnarray}
where the function $f_\nu(x)=2M\varepsilon_0\big(2x+2\nu+3\big)$.
By setting
\begin{eqnarray}
  {_\gamma}a_\nu:=\,{_\gamma}A_{\nu}\,\sqrt{\frac{N}{\mathcal {E}^{(\nu)} (N)}}\,e^{-i\gamma(\mathcal {E}^{(\nu)} (N)-\mathcal {E}^{(\nu)} (N-1))},\;
{_\gamma}a_\nu^\dag:=e^{i\gamma(\mathcal {E}^{(\nu)} (N)-\mathcal {E}^{(\nu)} (N-1))}\,\sqrt{\frac{N}{\mathcal {E}^{(\nu)} (N)}} \,{_\gamma}A_{\nu}^\dag,\cr
\end{eqnarray}
 the  algebra (\ref{sama:algebra}) becomes the 
Weyl-Heisenberg algebra, i.e.,
\begin{eqnarray}
  {_\gamma}a_\nu\, {_\gamma}a_\nu^\dag- 
\,{_\gamma}a_\nu^\dag\,{_\gamma}a_\nu ={\bf 1},\quad [N,\,{_\gamma}a_\nu]=-
\,{_\gamma}a_\nu,\quad [N, \,{_\gamma}a_\nu^\dag]= \,{_\gamma}a_\nu^\dag,\quad {_\gamma}a_\nu^\dag\,{_\gamma}a_\nu:=N.
\end{eqnarray}
\begin{definition}
The  GKCSs  associated with the 
annihilation operator (\ref{sama:eigen22}) are defined as follows \cite{KinaniD,gk,fbaga,DeyFring,Alietal}
\begin{eqnarray}
\label{sama:sama}
|z,\gamma\rangle_{\nu}:=\mathcal{N}_{\nu}^{-1/2}(|z|^2)
\sum_{n=0}^\infty\frac{z^n e^{-i\gamma
\mathcal {E}^{(\nu)}(n)}}{\sqrt{\rho_n}}|n\rangle_{\nu},\quad z\in D_R,
\end{eqnarray}
where the normalization 
factor $\mathcal{N}_{\nu}(x)$ is  given by
\begin{eqnarray}
\label{netsa}
  \mathcal{N}_{\nu}(x):=\sum_{n=0}^{\infty}\frac{1}{(2\nu+3)_n}\frac{\big(x(2M\varepsilon_0)^{-1}\big)^n}{n!}
=
\frac{\Gamma(2\nu+3)}{(x/2M\varepsilon_0)^{2\nu+2}}I_{2\nu+2}\left(\frac{x}{M\varepsilon_0}\right),\quad x=|z|^2
\end{eqnarray}
with
\begin{eqnarray}
\rho_n:=\mathcal {E}^{(\nu)}(n)!,\quad 
\mathcal {E}^{(\nu)}(0)!=1,\quad 
D_R=\{z\,\in\,\mathbb{C}:\,|z|<R\}.
\end{eqnarray}
\end{definition}
$R=\limsup_{n\to\infty} \sqrt [n]\rho_n$ is the radius of 
convergence of the series (\ref{netsa}) and
   $I_{\nu}(x)$ is the modified Bessel function of order $\nu$ (for more details, see \cite{magnus}).
The 
GKCSs (\ref{sama:sama}) can be re-expressed in the following form
\begin{eqnarray}
\label{sama:GKCSs}
 |z,\gamma\rangle_{\nu}=\frac{|z|^{2\nu+2}}{(2M\varepsilon_0)^{\nu+1}\sqrt{I_{2\nu+2}(|z|^2/M\varepsilon_0)}}
\sum_{n=0}^\infty\frac{ e^{-i\gamma
\mathcal {E}^{(\nu)}(n)}}{\sqrt{\Gamma(n+1)\Gamma(2\nu+3+n)}}\left(\frac{z}{2M\varepsilon_0}\right)^n|n\rangle_{\nu}.
\end{eqnarray}
We now aim at showing that the coherent states (\ref{sama:GKCSs}) satisfy the Klauder's criteria \cite{klaud,klaud2}.
To this end let us first
 prove the following lemma.
\begin{lemma}
\label{samalemma}
\begin{eqnarray}
\label{sama:le}
  \int_{c-\beta-i\infty}^{c-\beta+i\infty}\Gamma(\beta+x)\Gamma(x)a^{-x}dx=
 4i\pi a^{\beta/2}K_{\beta}(2a^{1/2}),\quad  Re (\beta) < c, \;a\in\mathbb{R}
\end{eqnarray}
 where $K_m(x)$  is the modified Bessel function of the second kind \cite{magnus}
\begin{eqnarray}
K_{m}(x) = \frac{\pi}{2}\, \frac{I_{-m}(x) - I_{m}(x)}{\sin{(m\pi)}}
\end{eqnarray}
and $I_{m}(|z|)$ is the modified Bessel function of the first kind \cite{magnus}  
\begin{eqnarray}
I_{m}(2|z|) = \sum_{n=0}^{+\infty}\frac{|z|^{n+m}}{\Gamma(n+1)\Gamma(n+m+1)}.
\end{eqnarray}
\end{lemma}
{\bf Proof.}
From the  formula \cite{grad} (see eq. EH II 83(34) on page 685)  for  $z=i\sqrt t$, we have 
\begin{eqnarray}
\int_{-c-i\infty}^{-c+i\infty}\Gamma(-\beta-s)\Gamma(-s)\left(\frac{t}{4}\right)^{\beta/2+s}ds=-
2\pi^2\,e^{i\pi\beta/2}H_\beta^{(1)}(it^{1/2}).
\end{eqnarray}
By setting $-s=x+\beta$ and $a=t/4$,  the latter formula becomes
\begin{eqnarray}
\int_{c-\beta-i\infty}^{c-\beta+i\infty}\Gamma(\beta+x)\Gamma(x)a^{-x}dx=-
2\pi^2\,e^{i\pi\beta/2}a^{\beta/2}H_{\beta}^{(1)}(2ia^{1/2}).
\end{eqnarray}
The proof is achieved by replacing $H_{\beta}^{(1)}(it)=\frac{2}{\pi}i^{-\beta-1}K_\beta(t).$
\hfill$\square$ 

\begin{proposition}
The GKCSs defined in (\ref{sama:GKCSs}) 
\begin{enumerate}
\item are not orthogonal to each other, i.e.,
\begin{eqnarray}
{}_\nu\langle z',\gamma|z,\gamma\rangle_{\nu}\neq \delta(z-z'),
\end{eqnarray}
\item are normalized,
\begin{eqnarray}
{}_\nu\langle z,\gamma|z,\gamma\rangle_{\nu}=1,
\end{eqnarray}
\item are continuous in their labels, i.e.,
\begin{eqnarray}
\forall\,z,z'\in \mathbb{C},\,\;\left\| \vert z,\gamma\rangle_{\nu} - \vert
z',\gamma\rangle_{\nu}\right\|^2 \to  0  \mbox{ as } |z-z'|\to 0,
\end{eqnarray}
\item solve the unity, i.e.,
\begin{eqnarray}
\label{sama:solv}
\int_{\mathbb{C}}d\mu_\nu(|z|^2)|z,\gamma
\rangle_{\nu}{}_\nu\langle z,\gamma|={\bf 1},
\end{eqnarray}
where the measure $d\mu_\nu(|z|^2)$ is given by
\begin{eqnarray}
  d\mu_\nu(|z|^2)=\frac{ (2M\varepsilon_0)^{\nu}}{\;|z|^{2\nu+2}}I_{2\nu+2}\left(\frac{|z|^2}{M\varepsilon_0}\right)K_{2\nu+2}\left(|z|\sqrt{\frac{2}{M\varepsilon_0}}\right)\,\frac{d^2 z}{\pi}.
\end{eqnarray}
\item are temporarily stable, i.e.,
\begin{eqnarray}
e^{-it{\cal H}_\nu}|z,\gamma\rangle_{\nu}=
|z,\gamma+t\rangle_{\nu}\quad t\,\in\,\mathbb{R},
\end{eqnarray}
\item satisfy  the action  identity, i.e.,
\begin{eqnarray}
{}_\nu\langle \gamma,z|{\cal H}_\nu|z,\gamma\rangle_{\nu}=
|z|^2.
\end{eqnarray}
\end{enumerate}
\end{proposition}
{\bf Proof.}

$\bullet$ The product of two GKCSs 
$|z,\gamma\rangle_{\nu}$ and $|z',\gamma
\rangle_{\nu}$ is given by
\begin{eqnarray}
\label{orthog}
 {}_\nu\langle z',\gamma|z,\gamma
\rangle_{\nu}&=&\left[\mathcal{N}_{\nu}(|z|^2)\mathcal{N}_{\nu}(|z'|^2)\right]^{-1/2}\sum_{n=0}^{\infty}\frac{((2M\varepsilon_0)^{-1}z\bar{z'})^n}{(2\nu+3)_n\,n!}\cr
&=&\left(\frac{|zz'|}{z\bar{z'}}\right)^{2\nu+2}\frac{I_{2\nu+2}(z\bar{z'} /M\varepsilon_0)}{\sqrt{I_{2\nu+2}(|z|^2/M\varepsilon_0)I_{2\nu+2}(|z'|^2/M\varepsilon_0)}}\cr
&\neq& \delta(z-z').
\end{eqnarray}
Therefore,  the GKCSs (\ref{sama:sama})  are not 
orthogonal but for $z'=z$, the product
(\ref{orthog}) is equal to $1$. 

$\bullet$
\begin{eqnarray}
\left\| \vert z,\gamma\rangle_{\nu} - \vert
z',\gamma\rangle_{\nu}\right\|^2 = 2\left(1 -  
Re(_{\nu}\langle z',\gamma\vert z,\gamma\rangle_{\nu})\right).
\end{eqnarray}
So,
$
\left\| \vert z,\gamma\rangle_{\nu} - \vert
z',\gamma\rangle_{\nu}\right\|^2 \to  0$ as $|z-z'|\to 0$, 
since $_{\nu}\langle z',\gamma\vert z,
\gamma\rangle_{\nu}\to 1$ as $\vert z - z'\vert \to   0.
$ The GKCSs (\ref{sama:sama}) are continuous in their labels.

$\bullet$ To investigate the resolution 
of unity  (\ref{sama:solv}), we assume the
existence of a positive weight 
$\mu_\nu(|z|^2)$ such that the resolution of the identity
reads
\begin{eqnarray}
 \int_{\mathbb{C}}d\mu_\nu\,(|z|^2)|z,\gamma\rangle_{\nu}
{}_\nu\langle z,\gamma|=\sum_{n=0}^\infty
|n\rangle_\nu{}_\nu\langle n|={\bf 1}.
\end{eqnarray}

Upon passing to polar coordinates,  
$z= re^{i\theta}$, $d\mu_\nu(|z|^2)= \omega_\nu (|z|^2)d^2z,\;d^2z=d(Re z)d(Im z)$
 and integrating with respect to
$\theta$, the  function $\omega_\nu (x)$  is required to be in the form
\begin{eqnarray}   
  \omega_\nu(x) =\frac{1}{\pi}{\tilde \omega}_\nu(x){\cal N}_\nu(x)  ,\quad x=r^2,
\end{eqnarray}
where the function ${\tilde \omega}_\nu(x)$ is to be determined from the
equation
\begin{eqnarray}
 \label{KaMoment}
 \int_0^{\infty}x^{n}
{\tilde \omega}_\nu(x)\,dx=
 \rho_n,\quad n= 0,\; 1,\; 2,\; \cdots.
\end{eqnarray}
If $n$ in (\ref{KaMoment}) is extended 
to $s-1$, where $s\in\mathbb{C}$, then 
the problem can be
formulated in terms of the Mellin and inverse Mellin 
transforms \cite{Jacqueline} that have
been extensively used in the context of 
various kinds of generalized CSs. By setting
$\rho_\nu(n)=\rho_n$, here $\rho_\nu(s-1)$ is the Mellin 
transform $\mathcal{M}[{\tilde \omega}_\nu(x);s]$ of
${\tilde \omega}_\nu(x)$, i.e.,
\begin{eqnarray}
\rho_\nu(s-1)=\mathcal{M}[{\tilde \omega}_\nu(x);s]\equiv\int_0^{\infty}x^{s-1}
{\tilde \omega}_\nu(x)\,dx,
\end{eqnarray}
and ${\tilde \omega}_\nu(x)$  is the inverse Mellin transform
$\mathcal{M}^{-1}[\rho_\nu(s-1);x]$ of $\rho_\nu(s-1)$, i.e.,
\begin{eqnarray}
  {\tilde \omega}_\nu(x)=\mathcal{M}^{-1}[\rho_\nu(s-1);x]&\equiv& \frac{1}{2i\pi}\int_{c'-i\infty}^{c'+i\infty}x^{-s}
\rho_\nu(s-1)\,ds\cr
%&=& \frac{1}{2i\pi}\int_{c'-i\infty}^{c'+i\infty}x^{-s}
%(2M\varepsilon_0)^{s-1}(s-1)!\frac{(2\nu+s+1)!}{(2\nu+2)!}\,ds\cr
&=&\frac{1}{4i\pi M\varepsilon_0\Gamma(2\nu+3)}\int_{c'-i\infty}^{c'+i\infty}a^{-s}
\Gamma(s)\Gamma(2\nu+2+s)\,ds,
\end{eqnarray}
where $a=x/2M\varepsilon_0$. By using the 
Lemma \ref{samalemma} 
for $\beta=2\nu+2$, we deduce that
\begin{eqnarray}
{\tilde \omega}_\nu(x)=2\frac{x^{\nu+1}}{(2M\varepsilon_0)^{\nu+2}
\Gamma(2\nu+3)}K_{2\nu+2}\big((2x/M\varepsilon_0)^{1/2}\big)
\end{eqnarray}
and
\begin{eqnarray}
d\omega_\nu(x)=\frac{2 (2M\varepsilon_0)^{\nu}}{\pi \;x^{\nu+1}}I_{2\nu+2}(x/M\varepsilon_0)K_{2\nu+2}\big(\sqrt{2x/M\varepsilon_0}\big)\,dx.
\end{eqnarray}

$\bullet$
Since ${\cal H}_\nu|n\rangle_{\nu}={\cal E}^{(\nu)}(n)|n\rangle_{\nu},$ the stability and action identity are obviously
true.
\hfill$\square$
%%%%%%%%%%%%%%%%%%%%%%%ù
\section{Statistics and geometry of the GKCSs $|z,\gamma\rangle_\nu$}
The conventional boson operators ${_\gamma}a_\nu$ and ${_\gamma}a_\nu^\dag$
have their actions  on the states $|n\rangle_\nu$  given by
\begin{eqnarray}
 {_\gamma}a_\nu|n\rangle_\nu=\sqrt{n}\,|n-1\rangle_\nu\quad \mbox{and} \quad 
{_\gamma}a_\nu^\dag|n\rangle_\nu=\sqrt{n+1}\,|n+1\rangle_\nu.
\end{eqnarray}
Besides,
\begin{eqnarray}
 ({_\gamma}a_\nu)^r|n\rangle_\nu=\sqrt{\frac{n!}{(n-r)!}}\,|n-r\rangle_\nu, \quad 0\leq r\leq n
\end{eqnarray}
and
\begin{eqnarray}
 ({_\gamma}a_\nu^\dag)^s|n\rangle_\nu=\sqrt{\frac{(n+s)!}{n!}}\,|n+s\rangle_\nu.
\end{eqnarray}
%%%%%%%%%%%%%%%%%%%%%%%%%
\subsection{ Statistics of the GKCSs $|z,\gamma\rangle_\nu$}
\begin{proposition}
 The expectation value of $({_\gamma}a_\nu^\dag)^s {_\gamma}a_\nu^r$ in the coherent states $|z,\gamma\rangle_\nu$ is given by
\begin{eqnarray}
\label{sama:abel:ma}
 \langle({_\gamma}a_\nu^\dag)^s \,{_\gamma}a_\nu^r\rangle
=\bar z^sz^r{\cal S}_{\nu,\gamma}^{(s,r)}(|z|^2),\quad s,r=0,\;1,\;2,\cdots
\end{eqnarray}
where
\begin{eqnarray}
\label{abel:saa}
{\cal S}_{\nu,\gamma}^{(s,r)}(x)=\frac{1}{\mathcal{N}_\nu(x)}\sum_{m=0}^\infty e^{i\gamma(\mathcal{E}^{(\nu)}(m+s)-\mathcal{E}^{(\nu)}(m+r))}\sqrt{\frac{(m+r)!(m+s)!}{\rho_{m+s}\rho_{m+r}}}\,\frac{x^m}{m!}.
\end{eqnarray}
In particular,
\begin{eqnarray}
 \langle({_\gamma}a_\nu^\dag)^r \,{_\gamma}a_\nu^r\rangle= \frac{x^{r}}{\mathcal{N}_{\nu}(x)}\left(\frac{d}{dx}\right)^r\mathcal{N}_{\nu}(x),\quad x=|z|^2,\quad r=0,\;1,\;2,\cdots,
\end{eqnarray}
and
\begin{eqnarray}
 \langle N\rangle= x\frac{\mathcal{N}_{\nu}'(x)}{\mathcal{N}_{\nu}(x)}\;,
\end{eqnarray}
where $()'$ denotes the derivative with respect to $x$.
\end{proposition}
{\bf Proof.}  Indeed, for $s=0,\;1,\;2,\cdots$ and $r=0,\;1,\;2,\cdots$, we have 
\begin{eqnarray}
 \langle({_\gamma}a_\nu^\dag)^s \,{_\gamma}a_\nu^r\rangle:&=&{}_{\nu}\langle z,\gamma|({_\gamma}a_\nu^\dag)^s \,{_\gamma}a_\nu^r|z,\gamma\rangle_{\nu}
\cr
&=&\frac{1}{\mathcal{N}_\nu(|z|^2)}\sum_{m=0}^\infty\sum_{n=r}^\infty e^{i\gamma(\mathcal{E}^{(\nu)}(m)-\mathcal{E}^{(\nu)}(n))}\sqrt{\frac{n!(n-r+s)!}{\rho_{m}\rho_{n}(n-r)!(n-r)!}}\bar z^mz^n{}_\nu\langle m|n+s-r\rangle_\nu
\cr
&=&\frac{1}{\mathcal{N}_\nu(|z|^2)}\sum_{n=r}^\infty 
e^{i\gamma(\mathcal{E}^{(\nu)}(n-r+s)-\mathcal{E}^{(\nu)}(n))}\sqrt{\frac{n!(n-r+s)!}{\rho_{n+s-r}\rho_{n}(n-r)!(n-r)!}}\bar z^{n+s-r}z^n
\cr
&=&\frac{\bar z^sz^r}{\mathcal{N}_\nu(|z|^2)}\sum_{n=0}^\infty e^{i\gamma(\mathcal{E}^{(\nu)}(n+s)-\mathcal{E}^{(\nu)}(n+r))}\sqrt{\frac{(n+r)!(n+s)!}{\rho_{n+s}\rho_{n+r}}}\frac{|z|^{2n}}{n!},
\end{eqnarray}

In the special case $s=r$, we have 
\begin{eqnarray*}
 \langle({_\gamma}a_\nu^\dag )^r\,{_\gamma}a_\nu^r\rangle&=& \frac{x^{r}}{\mathcal{N}_\nu(x)}\sum_{m=0}^\infty \frac{(m+r)!}{\rho_{m+r}}
\frac{x^{m}}{m!}
\cr&=& \frac{x^{r}}{\mathcal{N}_\nu(x)}\sum_{m=r}^\infty \frac{m!}{\rho_{m}}\frac{x^{m-r}}{(m-r)!}
\cr&=& \frac{x^{r}}{\mathcal{N}_\nu(x)}\left(\frac{d}{dx}\right)^r\mathcal{N}_\nu(x),\quad x=|z|^2.
\end{eqnarray*}
In particular, for $r=1$ the latter expression takes the form
\begin{eqnarray}
 \langle N\rangle\equiv\langle {_\gamma}a_\nu^\dag  \,{_\gamma}a_\nu\rangle= x\frac{\mathcal{N}_\nu'(x)}{\mathcal{N}_\nu(x)}.
\end{eqnarray}
\hfill$\square$

The probability of finding $n$ quanta in the deformed state $|z,\gamma\rangle_{\nu}$ is given by
\begin{eqnarray}
 \mathcal{P}_{\nu}(x,n):=|_\nu\langle n|z,\gamma\rangle_\nu|^2= \frac{x^n}{
{\cal E}^{(\nu)}(n)!\, \mathcal{N}_\nu(x)},\quad x=|z|^2.
\end{eqnarray}
Since for the non-deformed CS the variance of the number operator $N$ is equal to
its average, deviations from Poisson distribution can be measured with the Mandel
 parameter  defined by the quantity \cite{DeyFring}
\begin{eqnarray}
\label{sama:mand}
 \mathcal{Q}_{\nu} :=\frac{(\Delta N)^2-\langle N\rangle}{\langle N\rangle}
\end{eqnarray}
where $\langle N \rangle$ is the average counting number, $ (\Delta N)^{2}=\langle N^2\rangle-\langle N\rangle^2$ is the corresponding  square  variance.
Moreover, the Mandel parameter $ \mathcal{Q}_{\nu}$
\begin{eqnarray}
 \mathcal{Q}_{\nu}   \equiv \mathcal F - 1
\end{eqnarray}
 is closely related to the normalized variance,  also called the quantum 
Fano factor $\mathcal F$\; \cite{issiak,bajer-miranowicz}, 
  given   by $\mathcal F =  (\Delta N)^{2} /\langle N \rangle$, of 
the photon distribution. For 
$\mathcal F < 1 \,( \mathcal{Q}_{\nu} < 0)$, the emitted light 
  is referred to as  sub-Poissonian,   $\mathcal F = 1,  
\mathcal{Q}_{\nu} = 0$ corresponds to  the Poisson distribution while 
 for $\mathcal F > 1, ( \mathcal{Q}_{\nu} > 0)$ it corresponds to  
 super-Poissonian \cite{mandel1,zhangetal,jpjpg}.  

By using 
the expectation value of the operator  $N^2=( {_\gamma}a_\nu^\dag )^2\,{_\gamma}a_\nu^2+ N$ provided by
\begin{eqnarray}
 \langle N^2\rangle
= x^2\,{\cal S}_{\nu,\gamma}^{(2,2)}(x)+
%\frac{\mathcal{N}_\nu''(x)}{\mathcal{N}_\nu(x)}+
x\,{\cal S}_{\nu,\gamma}^{(1,1)}(x),\quad x=|z|^2
%\frac{\mathcal{N}_\nu'(x)}{\mathcal{N}_\nu(x)},
\end{eqnarray}
one readily finds
\begin{eqnarray}
 \mathcal{Q}_\nu(x)=x\left(\frac{\mathcal{N}_\nu''(x)}{\mathcal{N}_\nu'(x)} -\frac{\mathcal{N}_\nu'(x)}{\mathcal{N}_\nu(x)}\right).
\end{eqnarray}
For  $x<<1$, the mandel parameter 
(\ref{sama:mand}) is reduced
to
\begin{eqnarray}
\label{sama:poissonian}
 \mathcal{Q}_\nu(x)=-\frac{x}{2 M\varepsilon_0(2\nu+3)(2\nu+4)}+ o(x^2) < 0
\end{eqnarray}
which yields the sub-Poissonian distribution.

The second order correlation function
 defined as
\begin{eqnarray}
g_\nu^{(2)}(x):=\frac{ \langle N^2\rangle - \langle N \rangle }{ \langle N \rangle^2}, \quad x=|z|^2
\end{eqnarray}
is explicitly given by
\begin{eqnarray}
\label{samasama}
g_\nu^{(2)}(x)=\frac{\mathcal{N}_\nu''(x)\mathcal{N}_\nu(x)}{[\mathcal{N}_\nu'(x)]^2}.
\end{eqnarray}
For  $x<<1$, the second order correlation function
(\ref{samasama}) is reduced
to
\begin{eqnarray}
g_\nu^{(2)}(x)=\frac{2\nu+3}{2\nu+4}\left(1+\frac{x}{ M\varepsilon_0(2\nu+3)(2\nu+4)(2\nu+5)}\right)+ o(x^2).
\end{eqnarray}

The Hermitian   operators $X_\nu$ and $P_\nu$ defined as follows
\begin{eqnarray}
\label{sama:denu}
X_\nu:=\frac{ {_\gamma}a_\nu^\dag+ {_\gamma}a_\nu}{\sqrt{2}},\quad P_\nu:=i\frac{ {_\gamma}a_\nu^\dag- {_\gamma}a_\nu}{\sqrt{2}}
\end{eqnarray}
lead to the following uncertainty relation
\begin{eqnarray}
(\Delta X_\nu)^2(\Delta P_\nu)^2\geq \frac{1}{4}|\langle[ X_\nu,\, P_\nu] \rangle|^2.
\end{eqnarray}
When the expectation values are evaluated in the CSs
$|z,\gamma\rangle_\nu$ and we find that $\sigma_{X_\nu}<
\Delta_{{\cal H}_\nu}<\sigma_{P_\nu}$ ($\sigma_{P_\nu}<
\Delta_{{\cal H}_\nu}<\sigma_{X_\nu}$), we say that
$|z,\gamma\rangle_\nu$ is an $X_\nu-$squeezed 
state ($P_\nu-$squeezed state) \cite{Aleixo}.

Using the relation  (\ref{sama:abel:ma}),  the
variances of   the  operators
 $X_\nu$ and $P_\nu$ are evaluated in the 
state $|z,\gamma\rangle_\nu$ as
\begin{eqnarray}
\label{samasigma1}
  \sigma_{X_\nu}(z)&=&Re\left[\bar{z}^2({\cal S}_{\nu,\gamma}^{(2,0)}(|z|^2)-({\cal S}_{\nu,\gamma}^{(1,0)}(|z|^2))^2)\right]
\cr
&+&|z|^2\left({\cal S}_{\nu,\gamma}^{(1,1)}(|z|^2)-\left|{\cal S}_{\nu,\gamma}^{(1,0)}(|z|^2) \right|^2\right)+\frac{1}{2}
\end{eqnarray}
and
\begin{eqnarray}
\label{samasigma2}
  \sigma_{P_\nu}(z)&=&-Re\left[\bar{z}^2({\cal S}_{\nu,\gamma}^{(2,0)}(|z|^2)-({\cal S}_{\nu,\gamma}^{(1,0)}(|z|^2))^2)\right]
\cr
&+&|z|^2\left({\cal S}_{\nu,\gamma}^{(1,1)}(|z|^2)-\left|{\cal S}_{\nu,\gamma}^{(1,0)}(|z|^2) \right|^2\right)+\frac{1}{2},
\end{eqnarray}
where ${\cal S}_{\nu,\gamma}^{(s,r)}(x)$ is defined in (\ref{abel:saa}). From the relations (\ref{samasigma1}) and (\ref{samasigma2}), one can show that 
\begin{eqnarray}
\sigma_{P_\nu}(z)=\sigma_{X_\nu}(e^{i\frac{\pi}{2}}z).
\end{eqnarray}
 This relation means that to obtain the representation of $\sigma_{P_\nu}(z)$
in the same plane we only need to apply a positive rotation of $\frac{\pi}{2}$ to
 the $[Re(z), Im(z)]$-plane representation of $\sigma_{X_\nu}(z)$.

%%%%%%%%%%%%%%%%%%%%%%%%%%%
\subsection{ Geometry of the states $|z,\gamma\rangle_{\nu}$}
 The geometry of a quantum state space can be described by the corresponding metric tensor. This real and positive definite metric is defined on the underlying manifold that the quantum states form, or belong to, by calculating the distance function (line element) between
two quantum states. It is also known as a Fubini-Study metric of the ray space. The knowledge of the quantum metric enables  to calculate quantum mechanical transition probability and uncertainties
\cite{Dezi}.

The map  $z\longmapsto|z,\gamma\rangle_{\nu}$ defines a map from the space $\mathbb{C}$ of complex numbers onto
a continuous subset of unit vectors in Hilbert space and generates in the latter a two-dimensional surface with the following Fubini-Study metric:
\begin{eqnarray}
\label{sama:metric}
 d\sigma^2:= ||d|z,\gamma\rangle_{\nu}||^2-|_{\nu}\langle z,\gamma|d|z,\gamma\rangle_{\nu}|^2.
\end{eqnarray}
\begin{proposition}
The Fubini-Study metric  (\ref{sama:metric}) is reduced to
\begin{eqnarray}
 d\sigma^2= W_{\nu}(x)d\bar z dz,
\end{eqnarray}
where $x=|z|^2$ and 
\begin{eqnarray}
 W_{\nu}(x)=\left(x\frac{\mathcal{N}_{ \nu}'(x)}{\mathcal{N}_{\nu}(x)}\right)'= \frac{d}{dx}\langle N\rangle.
\end{eqnarray}
\end{proposition}
{\bf Proof.}
Computing $d|z,\gamma\rangle_{\nu}$ by taking into account the fact that any change of the form
$d|z,\gamma\rangle_{\nu}=\alpha|z,\gamma\rangle_{\nu}$, $\alpha\in\mathbb{C}$, has zero distance, we get
\begin{eqnarray}
 d|z,\gamma\rangle_{\nu}= \mathcal{N}_{\nu}^{-1/2}(|z|^2)\sum_{n=0}^\infty\frac{nz^{n-1}e^{-i\gamma\mathcal{E}^{(\nu)}(n)}}{\sqrt{\rho_{n}}}|n\rangle_{\nu}\;dz.
\end{eqnarray}
Then,
\begin{eqnarray}
 ||d|z,\gamma\rangle_{ \nu}||^2&=&\mathcal{N}_{ \nu}(|z|^2)^{-1}\sum_{n=0}^\infty\frac{n^2|z|^{2(n-1)}}{\rho_{n}}d\bar z dz
 \cr&=&\mathcal{N}_{ \nu}^{-1}(|z|^2)\left(\sum_{n=0}^\infty\frac{n|z|^{2(n-1)}}{\rho_{n}}
+|z|^2\sum_{n=0}^\infty\frac{n(n-1)|z|^{2(n-2)}}{\rho_{n}}\right)d\bar z dz
\cr&=&\mathcal{N}_{\nu}^{-1}(x)\left(\mathcal{N}_{\nu}'(x)+x\mathcal{N}_{\nu}''(x)\right)d\bar z dz
\cr&=&\mathcal{N}_{\nu}^{-1}(x)\left(x\mathcal{N}_{\nu}'(x)\right)'d\bar z dz
\end{eqnarray}
and
\begin{eqnarray}
 |_{ \nu}\langle z,\gamma|d|z,\gamma\rangle_{ \nu}|^2&=&
\left|\mathcal{N}_{ \nu}^{-1}(|z|^2)
\sum_{n=0}^\infty\frac{ n|z|^{2(n-1)}}{\rho_{n}}\bar z dz\right|^2
\cr&=&x\mathcal{N}_{ \nu}^{-2}(x)\left(\mathcal{N}_{ \nu}'(x)
\right)^2d\bar z dz.
\end{eqnarray}
Therefore,
\begin{eqnarray}
 d\sigma^2&=&\left(\frac{\mathcal{N}_{\nu}'(x)+x\mathcal{N}_{\nu}''(x)}{\mathcal{N}_{ \nu}(x)}-
x\left(\frac{\mathcal{N}_{\nu}'(x)}{\mathcal{N}_{\nu}(x)}\right)^2\right)d\bar{z}dz
\cr&=&\left(x\frac{\mathcal{N}_{\nu}'(x)}{\mathcal{N}_{\nu}(x)}\right)'d\bar{z}dz= \left(\frac{d}{dx}\langle N\rangle\right)d\bar{z}dz.
\end{eqnarray}
\hfill$\square$

For $x<<1$, we have
\begin{eqnarray}
 W_{\nu}(x)=\frac{1}{2M\varepsilon_0(2\nu+3)}\left(1-\frac{1}{M\varepsilon_0}\frac{x}{(2\nu+3)(2\nu+4)}\right)+o(x^2).
\end{eqnarray}
%%%%%%%%%%%%%
%%%%%%%%%%%%
\section{Quantization with the GKCSs}
The Berezin-Klauder-Toeplitz quantization, (also called ''anti-Wick" or coherent states quantization), of  phase space observables of the  complex plane, $D_R,$  uses the resolution of the identity (\ref{sama:solv}) and  is performed by 
mapping a function $f$ that satisfies appropriate conditions, to  the following operator in the Hilbert space  (see \cite{klauder63, gazeautt} and references therein for more details):
\begin{eqnarray}
\label{tota1}
  f\longmapsto A_f=\int_{D_R}f(z,\bar{z})|z,\gamma\rangle_{\nu}\, _{\nu}\langle z,\gamma|d\mu_\nu(|z|^2)=\sum_{n,n'=0}^\infty\left(A_f\right)_{nn'}|n\rangle_\nu{}_\nu\langle n'|,
\end{eqnarray}
where this integral is understood in the weak sense, i.e., it defines in fact a sesquilinear form (eventually only densely defined) 
\begin{eqnarray}
B_f(\psi_1,\psi_2)=\int_{D_R}f(z,\bar{z})\,\langle \psi_1|z,\gamma
\rangle_{\nu}\, _{\nu}\langle z,\gamma|\psi_2\rangle\,d\mu_\nu(|z|^2),
\end{eqnarray}
with the  matrix elements 
\begin{eqnarray}
\label{matr}
\left(A_f\right)_{nn'}=\frac{e^{i\gamma(\mathcal{ E}^{(\nu)}(n)-
\mathcal{ E}^{(\nu)}(n'))}}{\sqrt{\rho_n\rho_{n'}}}\int_{D_R}f(z,\bar{z})
z^{n}\bar{z}^{n'}\frac{d\mu_\nu(|z|^2)}{\mathcal{N}_\nu(|z|^2)}.
\end{eqnarray}
Operator $A_f$ is symmetric if $f(z,\bar{z})$ is real-valued,  and is 
bounded (resp. semi-bounded)
if $f(z,\bar{z})$  is bounded (resp. semi-bounded). In 
particular, the Friedrich extension allows
to define $A_f$ as a self-adjoint operator if $f(z,\bar{z})$ is 
a semi-bounded real-valued function.
Note that the original $f(z,\bar{z})$ is a "upper 
or contravariant symbol", usually non-unique, for the operator
$A_f.$ 
This problem involving the property of the function $f$ 
and the self-adjointness criteria of operators is thoroughly 
discussed  in a recent work by Bergeron et {\it al} \cite{berg} and 
does not deserve further development here.
So, without loss of generality, let us immediately examine 
different concrete expressions  for the function $f$ in the 
line of  \cite{gazeautt} as matter of result comparison:
\begin{enumerate}
 \item The function $f$  only depends on $|z|^2 =x:$  the 
matrix elements (\ref{matr}) take  the form 
\begin{eqnarray}
 \left(A_f\right)_{nn'}=\frac{2\,\delta_{n,n'}}{(2M\varepsilon_0)^{\nu+2}\Gamma(2\nu+3)\rho_n}\int_{0}^\infty  x^{n+1+\nu}f(x)K_{2\nu+2}\big((2x/M\varepsilon_0))^{1/2}\big)\,dx
\end{eqnarray}
\item The function $f$ only  depends on the angle $\theta = \arg z,$  
i.e., $f(z,\bar z) = F(\theta):$   the matrix elements (\ref{matr})  are given by 
\begin{eqnarray}
\label{matelang1}
 \left(A_f\right)_{nn'}= c_{n'-n}(F)\,  \frac{ e^{i\gamma(\mathcal{E}^{(\nu)}(n)-\mathcal{E}^{(\nu)}(n'))}(2M\varepsilon_0)^{\frac{n+n'}{2}}\Gamma(2\nu+3+\frac{n+n'}{2})\Gamma(1+\frac{n+n'}{2})}{\Gamma(2\nu+3)\;\sqrt{\rho_n\;\rho_{n'}}}
\end{eqnarray}
where  $c_n(F)$ are the Fourier coefficients of the function $F$   
\begin{eqnarray}
c_n(F) = \displaystyle \frac{1}{2\pi}\int_0^{2\pi} e^{-in\theta}\, F(\theta) d\theta
\end{eqnarray}
and the integral
\begin{eqnarray}
\label{sama:le}
\int_{0}^{\infty}dx \, x^{\mu}K_{\nu}(ax) = 2^{\mu - 1}a^{-\mu-1}\Gamma\left(\frac{1+\mu + \nu}{2}\right)\Gamma\left(\frac{1+\mu 
- \nu}{2}\right), \crcr
 \ [Re(\mu + 1 \pm \nu) > 0, Re (a)> 0 ].
\end{eqnarray}
 is used \cite{grad}.
\item The function $f(z,\bar{z})=z$ and $f(z,\bar{z})=\bar{z}:$  the operator
(\ref{tota1}) gives 
\begin{eqnarray}
\label{stoper1}
   A_z = \,_\gamma A_\nu, \; \quad A_{\bar z} =\,_\gamma A_\nu^{\dag}
\end{eqnarray}
which act on the states as $ | n \rangle_\nu$
\begin{eqnarray}
\label{stoper2}
&&_\gamma A_\nu \, | n \rangle_\nu =  \sqrt{\mathcal{E}_{n}^{(\nu)}} e^{i\gamma(\mathcal{E}^{(\nu)}(n)-\mathcal{E}^{(\nu)}(n-1))} | n-1\rangle_\nu\, ,
   \quad\; \,_\gamma A_\nu \, | 0 \rangle_\nu  = 0,\\
&& _\gamma A_\nu^{\dag} \, | n \rangle_\nu =
     \sqrt{\mathcal{E}_{n+1}^{(\nu)}} e^{-i\gamma(\mathcal{E}^{(\nu)}(n+1)-\mathcal{E}^{(\nu)}(n))}|n+1\rangle_\nu.
\end{eqnarray}
The state $|z,\gamma\rangle_{\nu}$ is eigenvector of $A_z = \,_\gamma A_\nu$ with eigenvalue $z$ 
like for standard CSs and
the operators $A_z$ and $A_{\bar{z}}$ satisfy the algebra (\ref{sama:algebra}), i.e
\begin{eqnarray}
[A_z,A_{\bar{z}}]=f_\nu(N),
\end{eqnarray}
as required.
\item  Let $f$  be the function defined as $f(z,\bar{z})=z^\alpha\bar{z}^\sigma,\;\alpha, \sigma\in\mathbb{N}\cup \{0\}:$ the matrix elements (\ref{matr}) of $A_f$ are given by
\begin{eqnarray}
\label{matelan}
 \left(A_f\right)_{nn'} &=&  \frac{ e^{i\gamma(\mathcal{E}^{(\nu)}(n)-\mathcal{E}^{(\nu)}(n'))}(2M\varepsilon_0)^{\frac{n+n'+\alpha+\sigma}{2}}}{\Gamma(2\nu+3)\;\sqrt{\rho_n\;\rho_{n'}}}\cr
 &&\times \;\Gamma\Big(2\nu+3+\frac{n+n'+\sigma+\alpha}{2}\Big)\Gamma\Big(1+\frac{n+n'+\sigma+\alpha}{2}\Big)\;\delta_{n-\sigma,n'-\alpha},
\end{eqnarray}
where (\ref{sama:le}) is used.
\end{enumerate}
 To end this section, let us turn back  for the cases where $f(z,\bar{z})=z$ and $f(z,\bar{z})=\bar{z}$. 
Interesting  results emerging from this context are  given by:
\begin{eqnarray}
_{\nu}\langle \gamma, z| A_z|z,\gamma\rangle_{\nu} = z, \qquad _{\nu}\langle\gamma, z|  A_{\bar z} |z,\gamma\rangle_{\nu} = \bar z,
\end{eqnarray}
\begin{eqnarray}
{}_\nu\langle\gamma, z|  A^{2}_z|z,\gamma\rangle_{\nu} = z^2, \quad 
_{\nu}\langle\gamma, z|  A^{2}_{\bar z} |z,\gamma\rangle_{\nu} = {\bar z}^2, \quad
 _{\nu}\langle \gamma,z|   A_{\bar z} A_z |z,\gamma\rangle_{\nu} = |z|^2, 
\end{eqnarray}
and
\begin{eqnarray}
 \;_{\nu}\langle\gamma, z|  A_z   A_{\bar z} |z,\gamma\rangle_{\nu} = |z|^2\left(1+4M\varepsilon_0\frac{\mathcal{N}_\nu'(|z|^2)}{\mathcal{N}_\nu(|z|^2)}\right) + 2M\varepsilon_0\big(2\nu+3\big).
\end{eqnarray}

%%%%%%%%%%%%%%%%%%%%%%%%%
%%%%%%%%%%%%%%%%%%%%%%%%%%%%
%
\section{Conclusion}
In this work, we have 
 constructed Gazeau-Klauder type coherent states for a  P\"oschl-Teller model. Relevant characteristics have been investigated.   Induced geometry and statistics 
have been    studied. Finally the coherent states quantization of the classical phase space observables has been presented and discussed.

\section*{Acknowledgements}
This work is partially supported by the Abdus Salam International
Centre for Theoretical Physics (ICTP, Trieste, Italy) through the
Office of External Activities (OEA) - \mbox{Prj-15}. The ICMPA
is also in partnership with
the Daniel Iagolnitzer Foundation (DIF), France.
%%%%%%%%%%%%
%%%%%%%%%%%%%%%
\section*{Appendix A. The normalization constant of the eigenvector $|\phi_n^{(\nu,\beta)}\rangle$}
By using the property of the eigenstates, we have
\begin{eqnarray}
\label{eqnart}
 &&\delta_{n,m} =: (\phi_{n}^{(\nu,\beta)},\phi_{m}^{(\nu,\beta)})\cr
&=&\overline{K_n^{(\nu,\beta)}}K_m^{(\nu,\beta)}\int_{0}^L dx \sin^{2\nu+n+m+2}\left(\frac{\pi x}{L}\right)\;
\exp\left\{-\frac{\beta \pi x}{L}\Big(\frac{1}{\nu+n+1}+\frac{1}{\nu+m+1}\Big)\right\}\cr
&&\times\;  \overline{P_n^{(a_n,\bar{a}_n)}\Big(i\cot \frac{\pi x}{L}\Big)}P_m^{(a_m,\bar{a}_m)}\Big(i\cot \frac{\pi x}{L}\Big)\\
 &=& \overline{K_n^{(\nu,\beta)}}K_m^{(\nu,\beta)}
\frac{(\bar{a}_n+1)_n(a_m+1)_n}{n!m!}\sum_{k=0}^m\sum_{s=0}^n  \frac{(-m,a_m+\bar{a}_m+m+1)_k(-n,a_n+\bar{a}_n+n+1)_s}{(\bar{a}_m+1)_k(a_n+1)_ss!k!}
{\bf \mathcal{J}} \nonumber
\end{eqnarray}
where 
\begin{eqnarray}
{\bf \mathcal{J}} &=&2^{-k-s}\int_{0}^L dx\Bigg[ \sin^{2\nu+m+n+2}\left(\frac{\pi x}{L}\right)
\exp\left\{-\frac{\beta \pi x}{L}\Big(\frac{1}{\nu+n+1}+\frac{1}{\nu+m+1}\Big)\right\}\cr
 &&\times \;\Big(1+i\cot \frac{\pi x}{L}\Big)^k
\Big(1-i\cot \frac{\pi x}{L}\Big)^s\Bigg]
\end{eqnarray}
In  \cite{berg} it is shown that  
\begin{eqnarray*}
 \int_{0}^1 dx \sin^{2\delta+2}(\pi x)e^{zx}=\frac{\Gamma(2\delta+3)e^{z/2}}{4^{\delta+1}\Gamma(\delta+2+i\frac{z}{2\pi})\Gamma(\delta+2-i\frac{z}{2\pi})}, \;\;n+\nu-\frac{k}{2}-\frac{s}{2}=\delta> -\frac{3}{2}.
\end{eqnarray*}
Therefore,
\begin{eqnarray}
\frac{\delta_{n,m}}{\overline{K_n^{(\nu,\beta)}}K_m^{(\nu,\beta)}}
&=&L\frac{(-\nu-m-\frac{i\beta}{\nu+m+1})_m(-\nu-n+\frac{i\beta}{\nu+n+1})_n}{n!m!\exp\big\{-\frac{\beta \pi }{2}\big(\frac{1}{\nu+n+1}+\frac{1}{\nu+m+1}\big)\big\}2^{2\nu+n+m+2}}\cr
&\times&\sum_{k=0}^m   \frac{(-m,-2\nu-m-1)_k}{(-\nu-m-\frac{i\beta}{\nu+m+1})_kk!\Gamma(\frac{n+m}{2}+\nu+2-k+\frac{i\beta}{\nu+n+1})}\cr
&\times&\sum_{s=0}^n \left\{\frac{(-n,-2\nu-n-1)_s\Gamma(n+m+2\nu-s-k+3)}{ \big(-\nu-n+i\frac{\beta }{2}\big(\frac{1}{\nu+n+1}+\frac{1}{\nu+m+1}\big)\big)_s s! }\right.\cr
&\times&\left.\frac{1}{\Gamma\big(\frac{n+m}{2}+\nu+2-s-i\frac{\beta }{2}\big(\frac{1}{\nu+n+1}+\frac{1}{\nu+m+1}\big)\big)}\right\}.
\end{eqnarray}
The proof is achieved by taking $n=m.$
%
%

% ****** End of file apssamp.tex ******
\end{document}